\documentclass[letterpaper, 10 pt, conference]{ieeeconf}  % Comment this line out if you need a4paper

\usepackage{authblk}

\usepackage{graphicx}
\usepackage{epsfig} % for postscript graphics files
\usepackage{fancyhdr}
\usepackage{stfloats}
\usepackage{amsmath} % assumes amsmath package installed
\usepackage{amssymb}  % assumes amsmath package installed
\usepackage{mathrsfs}
\usepackage{url}
\usepackage{tabularx,ragged2e,booktabs,caption}
\usepackage{diagbox}
\usepackage{bm}
\usepackage{bbm}
\usepackage{color, colortbl}
\usepackage[makeroom]{cancel}
\usepackage{theorem}
\usepackage{caption}
\usepackage{subcaption}
\usepackage{supertabular}
\usepackage{hyperref}
\usepackage{float}
\usepackage{longtable}
\usepackage{array} % for extrarowheight
\allowdisplaybreaks

\IEEEoverridecommandlockouts                              % This command is only needed if 
\title{\LARGE \bf
	Transformer Networks for Predictive Group Elevator Control
}

\author{Jing Zhang$^{1}$, Athanasios Tsiligkaridis$^{2}$, Hiroshi Taguchi$^{3}$, Arvind Raghunathan$^{1}$, and Daniel Nikovski$^{1}$% <-this % stops a space
	\thanks{$^{1}$Jing Zhang, Arvind Raghunathan, and Daniel Nikovski are with Mitsubishi Electric Research Laboratories (MERL), Cambridge, MA 02139, USA.
		{Email: \tt\small \{jingzhang, raghunathan, nikovsiki\}@merl.com}}%
	\thanks{$^{2}$Athanasios Tsiligkaridis is with the Department of Electrical and Computer Engineering, Boston University,
		Boston, MA 02215, USA; this work was done during his stay at MERL.
		{Email: \tt\small atsili@bu.edu}}%
		\thanks{$^{3}$Hiroshi Taguchi is with Advanced Technology R\&D Center, Mitsubishi Electric Corporation, Tokyo 100-8310, Japan.
		{Email: \tt\small Taguchi.Hiroshi@dw.MitsubishiElectric.co.jp}}%
}

\begin{document}

	\maketitle
	\thispagestyle{empty}
	\pagestyle{empty}

	%%%%%%%%%%%%%%%%%%%%%%%%%%%%%%%%%%%%%%%%%%%%%%%%%%%%%%%%%%%%%%%%%%%%%%%%%%%%%%%%
	\begin{abstract}
		We propose a Predictive Group Elevator Scheduler by using predictive information of passengers arrivals from a Transformer based destination predictor and a linear regression model that predicts remaining time to destinations. Through extensive empirical evaluation, we find that the savings of Average Waiting Time (AWT) could be as high as above 50\% for light arrival streams and around 15\% for medium arrival streams in afternoon down-peak traffic regimes. Such results can be obtained after carefully setting the Predicted Probability of Going to Elevator (PPGE) threshold, thus avoiding a majority of false predictions for people heading to the elevator, while achieving as high as 80\% of true predictive elevator landings as early as after having seen only 60\% of the whole trajectory of a passenger.
		
	\end{abstract}

	%%%%%%%%%%%%%%%%%%%%%%%%%%%%%%%%%%%%%%%%%%%%%%%%%%%%%%%%%%%%%%%%%%%%%%%%%%%%%%%%
	\section{INTRODUCTION}
	Group elevator control (GEC) is a demanding industrial control problem that needs to be solved repeatedly within a guaranteed time during the operation of a bank of several elevators in a building, under the conditions of very significant uncertainty stemming from various sources. The job of the group control system of an elevator bank is to decide how to transport vertically an endless stream of passengers who indicate their request for travel using call buttons located at the elevator landings and inside the elevator cars, by assigning passengers to one of the cars in the elevator bank. The objective is to optimize a suitable performance metric. most often to minimize the average waiting time (AWT) of passengers, but often other components can be used, such as the total travel time, consumed energy, etc.    
	
	Because the decision about the current passenger who is requesting service at the current time would affect the motion of the elevator cars in the future, GEC is not an instantaneous, but a sequential decision making problem with a potentially infinite optimization horizon. Even if the exact time of arrival of every passenger within a finite time interval was known, along with their arrival and destination floors, an exhaustive enumeration of all possible assignment schedules would lead to a combinatorial explosion, and is not a viable solution strategy. To make things worse, usually none of this information is known for passengers who are yet to arrive in the future. Instead, the GEC system usually knows only the arrival floor of the passengers who have already arrived, and either only their intended direction of travel (if traditional up/down hall-call buttons are installed at landings) or, possibly, the actual destination floor as well, if a full set of the newer destination-dispatch hall-call buttons are used \cite{Al-Sharif2015EstablishingBenchmarks:}.         
	
	Facing this lack of information about future arrivals of passengers, current GEC solutions used in industry usually make a radical simplification: ignore future arrivals altogether. Although such a simplification makes computation much easier, and allows the GEC system to meet real-time response requirements, it is clear that such a myopic solution strategy could not possibly be optimal. Attempts to consider future arrivals during the decision process have focused on exploiting the statistical properties of the stochastic arrival process, as well as analyzing the consequences of making a current assignment on future passengers. Pepyne and Cassandras  \cite{Pepyne1997OptimalTraffic} found analytically an optimal policy for pure up-peak elevator traffic that dispatches cars when their occupancy exceeds a threshold related to the arrival intensity of passengers, the performance characteristics of the elevators, as well as the number of cars available on the first floor. Nikovski and Brand \cite{Nikovski2003MarginalizingControl} used a semi-Markov chain with parameters (transition costs) derived from the arrival rate at a lobby floor to estimate the waiting times of both current and future passengers at the lobby resulting from various patterns of cars landing at that lobby, and use these estimates to select an assignment that would produce the optimal pattern. Empirical evaluation in mixed up-peak traffic (mostly from the lobby up, but also with some inter-floor traffic) demonstrated savings of waiting time with respect to a traditional myopic GEC algorithm on the order of $5\%$-$55\%$.
	
	As competitive as these methods can be with respect to traditional myopic GEC algorithms, they still use information about future arrivals only in its aggregate form, that is, they use the average arrival rate of passengers, with the assumption that it follows some standard arrival process, such as a Poisson one. This arrival rate can be estimated relatively easily from button presses at hall-call panels. However, even if the arrival process model is statistically correct in the aggregate, it does not specify precisely when individual passengers can be expected at landings. In contrast, recent advances in sensing and communication technologies have made it possible to collect much more detailed and accurate information about potential future arrivals at elevator landings, and deliver it in real time to the GEC system for the purposes of better decision making. Various technologies can be used to track the location and movement of building occupants in real time and communicate it to a central server, often originally for the purposes of security and access control. These technologies include RFID sensors attached to ID badges (\cite{Suzuki2013AInformation}), active badges with radio beacons, in-floor sensors (\cite{Orr2000TheTracking}), cameras that track occupants' movements, infrared motion detectors (\cite{Ivanov2007TrackingSystems}), as well as reflections of WiFi radio signals on the bodies of building occupants \cite{Adib20143DReflections}.
	
	How to best use this information for GEC is still an active area of research. Luh et al. formulated the problem of assigning all passengers within a time window of fixed duration (that also includes future passengers), and is updated in a rolling-horizon fashion, as a mixed-integer programming (MIP) problem, and proposed a solution algorithm based on Lagrangian relaxation, decomposition, and dynamic programming \cite{Luh2008GroupModes}. Kwon et al. (\cite{Kwon2014Sensor-awareEnvironments}) described a sensor-aware GEC method that places reservation calls for future passengers that are yet to arrive, assigns a car to service such reservation calls along with the other passengers already assigned to it, and potentially also moves that car proactively towards the arrival floor before the actual hall call has occurred.
	
	Although these methods for using advance arrival information can be very effective in reducing the waiting times of passengers, they assume that these arrivals would occur with complete certainty. This assumption can have detrimental effects on waiting times of passengers when it is not justified. Imagine, for example, a passenger who is likely to request elevator service with only $70\%$ probability. If a reservation call is made for such a passenger, a car is selected and dispatched to pick him/her up, and he/she ultimately does not request service ($30\%$ of the time), the car's trip would have been in vain. In addition to wasting energy to move the car, its movement might leave a large part of the building with no elevator cars in it, and passengers arriving in that part later might have to wait longer. Clearly, ignoring the uncertainty in passenger arrivals might actually end up being counterproductive.
	
	What is needed, then, is a method for accurately assessing the probability of passenger arrivals (both the arrival event and its time), as well as a GEC algorithm that can make good use of this uncertain information. The next section proposes one such algorithm, based on a predictive model for passenger requests for service that employs Transformer neural networks, and a predictive group elevator controller that makes use of the probabilistic predictions produced by the Transformer neural networks to  estimate  the expected waiting time resulting from possible assignments of future passengers. Section \ref{sec:empirical} presents an empirical evaluation of the predictive accuracy of Transformer networks in comparison to other methods, and analyzes, again empirically, the effect of the use of the predictive algorithm on reducing the AWT of passengers. 
	
	\section{PREDICTIVE GROUP ELEVATOR CONTROL WITH TRANSFORMER NETWORKS}
	The overall operation of the proposed predictive group elevator scheduler is organized as follows. A position tracking system located on each floor of a building tracks the sequence of positions of individual users over time, parses them into trajectory segments, each starting at an initial location and ending with a destination location, and stores them in a database of training data. This data is used to train a Transformer neural network to predict the final destination location based on partial trajectories. At run time, the trained network is used to continuously predict a probability distribution over all possible destinations for a specific person moving around the floor, using as input the partial trajectory registered for this particular person so far. The probability that this person will go to the elevator landing, along with the estimated time of arrival there, are passed on to the Predictive Group Elevator Scheduler (PGES). It integrates these predictions with similar information from all floors of the building to make optimal decisions about assigning elevator cars to actual requests for vertical transportation, as well as proactively moving the elevator cars even before such requests have been registered.

	\subsection{Overview and Principle of Operation of the PGES}
	
\begin{figure}[thpb]  
		\centering
 		\includegraphics[scale=0.26]{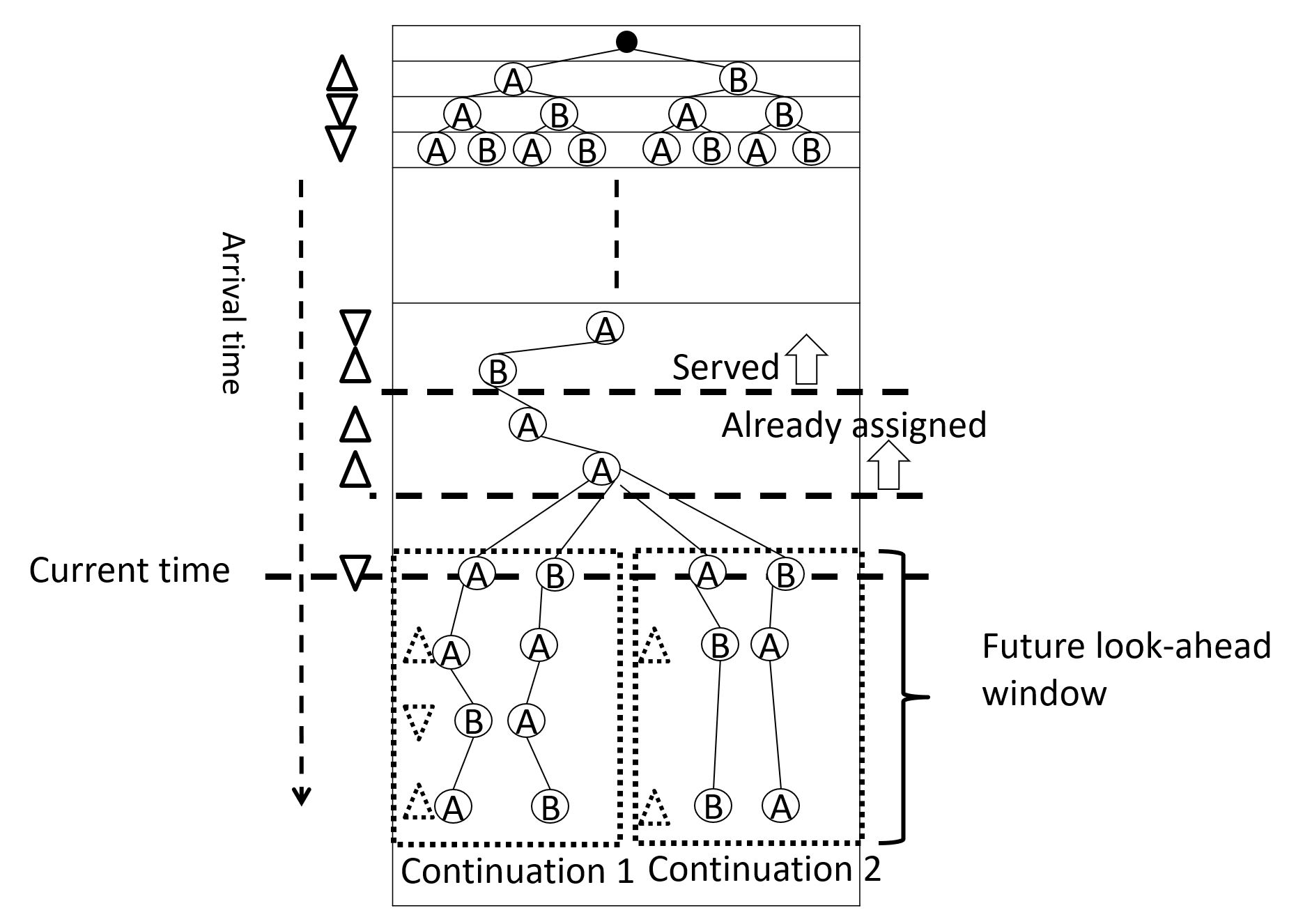}
		\caption{Operation of the predictive group elevator scheduler in immediate assignment mode. Predictions are used to generate one or more possible future continuations of the passenger arrival stream (not necessarily containing the same passengers), the currently arrived passenger is assigned tentatively to each car in turn (A or B in this case), and the AWT of all passengers, existing and future, for a given car assignment, is computed by forward simulation of the movement of all cars, and averaged over all continuations.}
		\label{fig:continuations-1}
	\end{figure}

\begin{figure}[thpb]  
		\centering
		\includegraphics[scale=0.33]{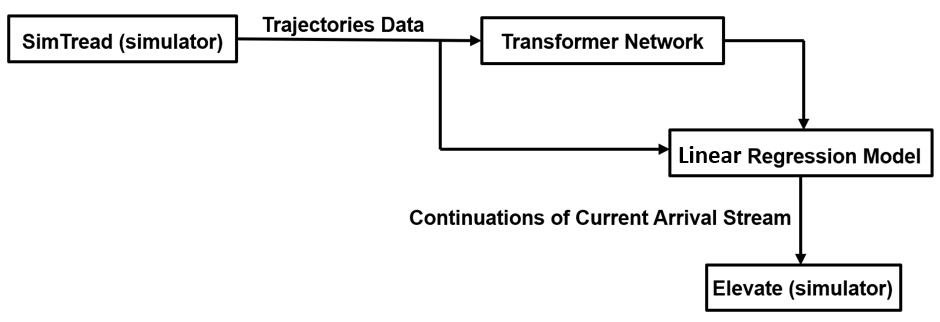}
		\caption{The structure of our PGES system.}
		\label{fig:simulators}
	\end{figure}

		The overall structure of our PGES system is shown in Fig. \ref{fig:simulators}. After obtaining sufficient training/testing trajectories data floor-by-floor (by using the SimTread simulator \cite{simtread}), we train a Transformer neural network for destination prediction and a linear regression model for the remaining time to destination prediction.  The Transformer network was first introduced in \cite{vaswani2017attention} as a sequence-to-sequence prediction model operating on discrete symbols, with superior predictive performance due to the attention mechanism it uses. In our previous work \cite{tsiligkaridis2020personalized}, we applied Transformer networks to predict passengers' destinations located on a building floor, based on previous partial trajectories of these passengers. We use these predictions for the purpose of GEC, when a decision is needed about which elevator car to assign to pick up a newly arrived passenger who has requested elevator service. At such a time, the Transformer network is executed in turn for each person who is currently in motion around each building floor, in order to determine whether this person is a likely future elevator passenger whose potential future request for service might affect (or be affected by) the currently registered request. To this end, the output of the Transformer network is interpreted probabilistically as a multinomial distribution over all possible destinations, including the elevator landing. This multinomial distribution is sampled to determine whether the passenger might be going to the elevator landing, and if yes, he/she is placed in a possible tentative future continuation of the current arrival stream for the elevator bank. Furthermore, if a person is determined to be a possible elevator passenger as a result of sampling, his/her arrival time at the elevator landing is predicted by means of a separately trained linear regression model. Depending on the amount of uncertainty in the multinomial distribution, one or more possible continuations are formed and used for determining the optimal car for the current passenger, by minimizing the expected Average Waiting Time (AWT) of both existing as well as tentative future passengers present in the continuations. In the general case, the expectation of AWT is computed by averaging over its value across multiple continuations, thus effectively taking a Monte-Carlo expectation over all multinomial distributions over future destinations produced by the Transformer network for each possible future passenger. Assuming immediate assignment mode, where a car is assigned to the newly arrived passenger at the time of his/her arrival, and never reconsidered later (as is usual in Japan and other countries), the computational time of this decision procedure is only linear in the number of available cars, because it is sufficient to tentatively assign the new passenger to each car in turn, evaluate the expected AWT of all passengers under this assignment as described above, and after that select the car resulting in the lowest expected AWT. The sequential operation of the scheduler in immediate assignment mode is shown in Fig. \ref{fig:continuations-1}, for two continuations, and further details about this procedure can be found in \cite{nikovski2017method}. The operation of the scheduler in immediate assignment mode essentially amounts to greedy search of the assignment tree, as shown in Fig. \ref{fig:continuations-1}, where each possible car assignment for a passenger is considered only once, at the time of arrival of that passenger, the best assignment is selected, and all others are ignored thereafter.

% 	\begin{figure}[thpb]  
% 		\centering
% 		\includegraphics[scale=0.26]{figs/continuations.png}
% 		\caption{A schematic of predictive group elevator scheduling with two continuation sets.}
% 		\label{fig:continuations}
% 	\end{figure}

	\subsection{Detailed Description of PGES}
	
	\subsubsection{Representation of Trajectories}    \label{sec:trajectories}
	Let us first describe how to represent trajectories to simplify analysis for a processor. Note that raw recordings from sensors for both positions and time stamps are continuous variables. It would be infeasible for a sequence-to-sequence predictor to directly use continuous values. Thus, we preprocess the positional and timing data by discretizing them into grid indices. 
	
	Let $x$'s and $y$'s denote longitudes and latitudes on a building floor, respectively.
	Arbitrarily take (but fix) lower bounds (resp., upper bounds) \(\underline x\) and \(\underline y\) (resp., \(\overline x\) and \(\overline y\)) for longitudes and latitudes respectively. Then the grid index \(p^\ast\) of any given tuple of coordinates \(\left( {{x},{y}} \right)\) in an $N_X \times N_Y$ rectangle can be determined as
	\(p^\ast = y^\ast N_X + {x}^\ast + 1\), where \({x}^\ast = \left\lfloor\frac{x - \underline x}{\Delta x}\right\rfloor\) and \({y}^\ast = \left\lfloor\frac{y - \underline y}{\Delta y}\right\rfloor\), with \emph{quantization intervals} being \(\Delta x = \frac{\overline x - \underline x}{N_X}\) and \(\Delta y = \frac{\overline y - \underline y}{N_Y}\). Note that the positive integers $N_X$ and $N_Y$ could be taken reasonably large, as long as the computational resources permit. Note also that $N_X$ and $N_Y$ could be determined by use of cross-validation, given a set of trajectory data.
	
	Given an upper bound $Q$ of \emph{remaining time to destination}, we determine the remaining time to destination for any recorded time stamp $t$ of a given tuple of positional coordinates \(\left( {{x},{y}} \right)\) in a given trajectory, whose last time stamp is recorded as $t_d$ (corresponding to the destination), by \({t^\ast} = t_d - t\). Taking both positional and timing information into account, a trajectory is denoted as 
	\begin{align}
	S = ((p^\ast_{1},t^\ast_{1}),(p^\ast_{2},t^\ast_{2}),\ldots,(p^\ast_{n},t^\ast_{n})),    \label{eq:traj}
	\end{align}
	where \(n\) is the length of the trajectory, and the last tuple \((p^\ast_{n},t^\ast_{n})\) might correspond to the destination, provided that the trajectory is complete.

	\subsubsection{Predictors}
	
	We build two predictors to deal with the positional and timing information separately -- one is based on the Transformer network, and the other is based on linear regression. To that end, we extract the positional sequence from \eqref{eq:traj} as 
	\begin{align}
	S_p = (p^\ast_{1},p^\ast_{2},\ldots,p^\ast_{n}),   \label{eq:traj-p}
	\end{align}
	which would be dealt with by the Transformer network.
	And, for each destination, we build a linear regression model $t^\ast = \beta_1 x^\ast + \beta_2 y^\ast$ ($\beta_1, \beta_2$ are two parameters to be fitted) to predict remaining time to the destination, given the current position index. Specifically, we solve the following optimization problem for a fixed destination:
	\begin{align}
	    \mathop {\min }\limits_{{\beta _1},{\beta _2} \in \mathbb{R}} \sum\limits_{i = 1}^m {{{\left( {{\Tilde{t}_i^\ast} - {\beta _1}{\Tilde{x}^\ast_{i}} - {\beta _2}{\Tilde{y}^\ast_{i}}} \right)}^2}},    \label{Lasso}
	\end{align}
	where $\{(\Tilde{x}_i^\ast, \Tilde{y}_i^\ast), \Tilde{t}_i^\ast, i = 1, \ldots, m\}$ are the training data. Assuming $(\beta_1^\ast, \beta_2^\ast)$ is the optimal solution to \eqref{Lasso}, then the resulting regression model would be $t^\ast = \beta_1^\ast x^\ast + \beta_2^\ast y^\ast$.
	Note that for the purpose of predicting people's remaining time to a destination, we have implicitly assumed people would always choose the shortest path to their destination and their walking speeds are the same; thus, given the current position, the remaining time to a selected destination would be uniquely determined. It is worth pointing out that in practice people's walking speeds are not necessarily the same, and in such cases, we can easily extend \eqref{Lasso} by adding the walking speed $v^\ast$ as an additional explanatory variable, thus leading to a regression model in the form of $t^\ast = \beta_1^\ast x^\ast + \beta_2^\ast y^\ast + \beta^\ast_3 v^\ast$. Admittedly, for the latter cases, we need more training data to obtain a satisfactorily accurate model.

	\subsubsection{Generation of Continuations}

When the uncertainty in the final destination of the current passenger is significant (that is, more than one destinations are likely, and having significant probability mass in the multinomial distribution), the previously described process of repeated sampling of this distribution to decide whether to place him/her in a continuation can be performed, and then followed by averaging the resulting AWT over multiple continuations to evaluate its expected value. However, when the multinomial distribution over continuations is significantly skewed towards one destination (likely the elevator landing for people who are going there, or another destination for people who are not), a simpler computational procedure can be followed to produce a single most-likely continuation.   The resulting continuation would then be utilized by the PGES directly, as if it was the single correct sequence of future arrivals.

The single continuation can be constructed as follows. Denote by $T$ the prediction length. Given a partial trajectory $S$ of a passenger, if the predicted probability (outputted by the Transformer network model) of the passenger going to the elevator is larger than some threshold $\delta$, we decide to place this person in the continuation, and otherwise ignore him/her. If included, we apply the linear regression model to predict the remaining time $t^\ast$ to the elevator and, if $t^\ast \le T$, we pass the predicted future arrival information of this passenger to the PGES (implemented in the simulator Elevate \cite{Elevate}). This use of a single continuation follows the certainty equivalence principle, and can be expected to be effective when the predictive method is fairly certain that the potential passenger is going to the elevator landing or not. A reasonable expectation is that the certainty of the predictive method will grow with time, as more tracking information is accumulated, and the passenger either approaches the elevator landing, or diverges from it en route to some other destination on the floor. This expectation is in fact supported by the empirical evaluation presented in the next section. Note that this method still uses probabilistic predictions, because it compares the probability that a potential passenger is going to the elevator against a threshold. How this threshold can be determined is explained in the next section.

\subsubsection{Destination Control Dispatcher}
	
We use a group control algorithm that uses information from a destination dispatch (DD) input panel that every passenger uses to register their destination floor. (The algorithm can easily be applied to the more common case of using only up/down hall-call panels, too, if the algorithm makes an assumption about the likely destination floor, as is customary in many GEC algorithms.) Each time a new
call is registered, the dispatcher assigns it tentatively in turn to each of the available cars, and evaluates the cost resulting from each tentative assignment, including the component of the cost for future passengers. The assignment leading to the lowest cost is then adopted. Suitable cost functions include
the average waiting time, the average journey time, or a combination of both. In this paper, we only use average waiting time. In particular, we implement the following DD algorithm (refer to \cite{barney2015elevator}) within the development interface of the commercial simulator Elevate \cite{Elevate}:
	
	Consider that a new call is to be allocated to an elevator bank of $C$ cars, each car ($c$) with $N(c)$ calls to answer and $WT(c)$ accumulated waiting time for the $N(c)$ calls. Let $NWK(c')$ denote the new accumulated waiting time for $N(c')+1$ calls, when the new call is allocated to car $c'$. Then the average waiting time for all calls is
	\begin{align}
	    AWT = \frac{NWT(c') + \sum^C_{c=1,c \ne c'}WT(c)}{1 + \sum^C_{c=1} N(c)},   \notag
	\end{align}
	which can be rewritten as
		\begin{align}
	    AWT = \frac{NWT(c') - WT(c')}{1 + \sum^C_{c=1} N(c)} + \frac{\sum^C_{c=1}WT(c)}{1 + \sum^C_{c=1} N(c)}.   \label{dd_awt}
	\end{align}
	From \eqref{dd_awt}, it is seen that minimizing $AWT$ is equivalent to minimizing $NWT(c') - WT(c')$, where $NWT(c')$ and $WT(c')$ can be evaluated by simulation.

	\section{EMPIRICAL EVALUATION}    \label{sec:empirical}

	\subsection{Experimental Setup}     \label{sec:building}

	\begin{figure}[thpb]  
		\centering
		\includegraphics[scale=0.3]{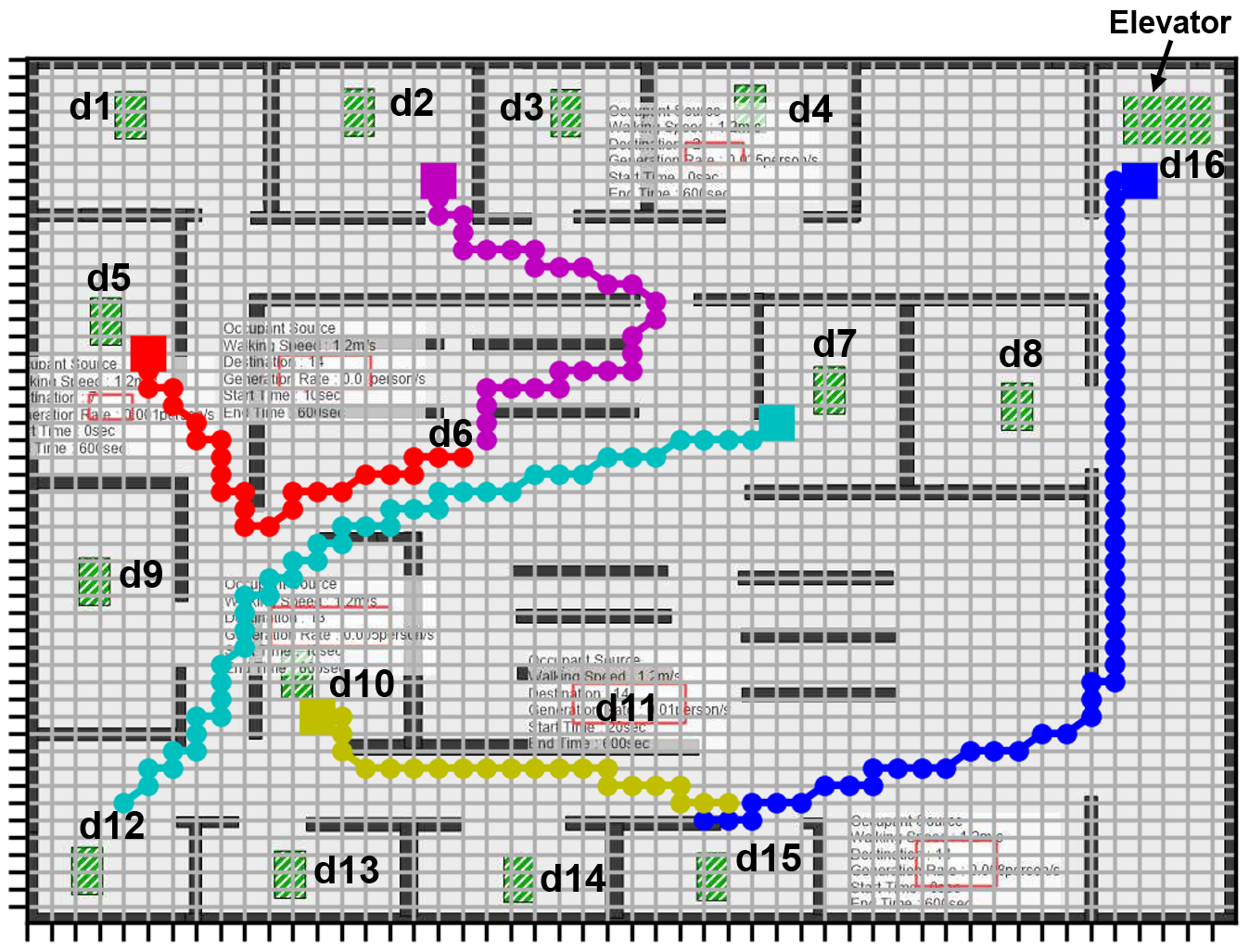}
		\caption{The layout of a floor in a simulated building.}
		\label{fig:merl_flr}
	\end{figure}

	We generate people's movement data in a building floor-by-floor. In particular, we use the SimTread software package for this purpose. The layout of a floor in a building that we use in the simulation is shown in Fig. \ref{fig:merl_flr}, where we have 16 destinations denoted d1 through d16, with d16 being the elevator. For simplicity, we consider an 8-floor building with floors 2 through 8 having exactly the same layout as shown in Fig. \ref{fig:merl_flr}; as traffic is a stochastic process, its realizations end up being different on each floor. Also shown in Fig. \ref{fig:merl_flr} are five typical trajectories; for example, the blue connected circles starting from d15 and ending at d16 represent a complete trajectory with the elevator as its destination, from which we can also see the evolution of position coordinates (formatted as a sequence of grid indices). Note that in our simulations, we assume all destinations d1 through d16 could also be an origin.
	To generate training/testing trajectories data, we use the prior probabilities specified in Table \ref{tab:prior_probs}.
	We consider a down-peak period of a typical workday; in particular, if a person heads to the elevator, then he/she is assumed to  go down to the lobby (the first floor) with the elevator, no matter which floor he/she currently is located at, and leave the building. Note that although the prior probability of going to the elevator ($0.2$ in this case) is higher at the end of the day than during other time periods, it is still not very high, so when a passenger starts moving, it is by far not a foregone conclusion that he/she will be going to the elevator. In order to pre-dispatch a car for this presumptive passenger, the GEC needs a much higher posterior probability, and it is the job of the predictor to produce it, for passengers that actually will need elevator service, after having observed the passenger's initial trajectory.

	Next, we first present detailed results for a fixed arrival rate, and then briefly show integrated results for various arrival rates.
	
	\begin{table*}
		\centering
		\caption{Prior probabilities for destinations.}
		\begin{tabular}{|c|c|c|c|c|c|c|c|c|c|c|c|c|c|c|c|c|}
			\hline
			Destination & d1 & d2 & d3 & d4 & d5 & d6 & d7 & d8 & d9 & d10 & d11 & d12 & d13 & d14 & d15 & d16 (elevator) \\
			\hline
			Probability & 0.05 & 0.07 & 0.02 & 0.07 & 0.05 & 0.07 & 0.16 & 0.02 & 0.05 & 0.1 & 0.05 & 0.02 & 0.02 & 0.05 & 0 & 0.2\\
			\hline
		\end{tabular}
		\label{tab:prior_probs}
	\end{table*}

	\subsection{Accuracy of the Transformer-Based Destination Predictor}
	
	Of particular interest is predicting whether a person will go to the elevator or not. In this section, we present results from the use of the Transformer-based destination predictor. For economy of space, we temporarily only show results for a single floor (floor $8$) from a single simulation scenario, where the arrival rate of passengers for the elevator is $64.8$ persons per hour (pph). 
	
	To build the Transformer-based destination predictor, we generate $55$ training trajectories for floor $\#8$, out of which $11$ have the elevator (d16) as the destination, covering a period of $10$ minutes. To generate test trajectories, we keep the same simulation parameters, but only change the random seed of the simulator, thus ending up with the same numbers of trajectories ($55$ in total, and $11$ of them have the elevator as the destination). When formatting the trajectories data, we take $N_X = N_Y = 50$ (refer to Section \ref{sec:trajectories}).
	
\begin{figure*}
\centering
\begin{subfigure}{.5\textwidth}
  \centering
  \includegraphics[width=.9\linewidth]{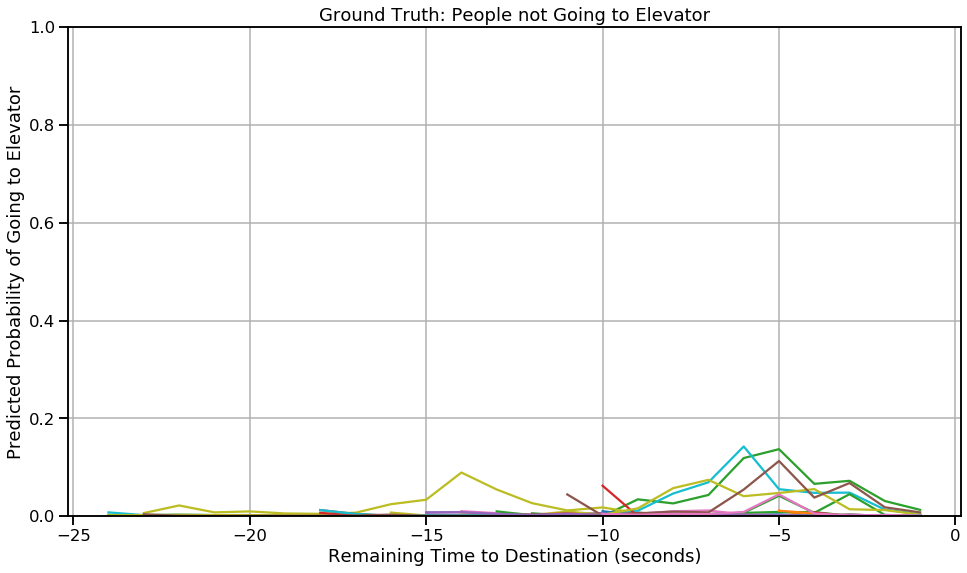}
  \caption{}
 	\label{fig:ppge2}
\end{subfigure}%
\begin{subfigure}{.5\textwidth}
  \centering
  \includegraphics[width=.9\linewidth]{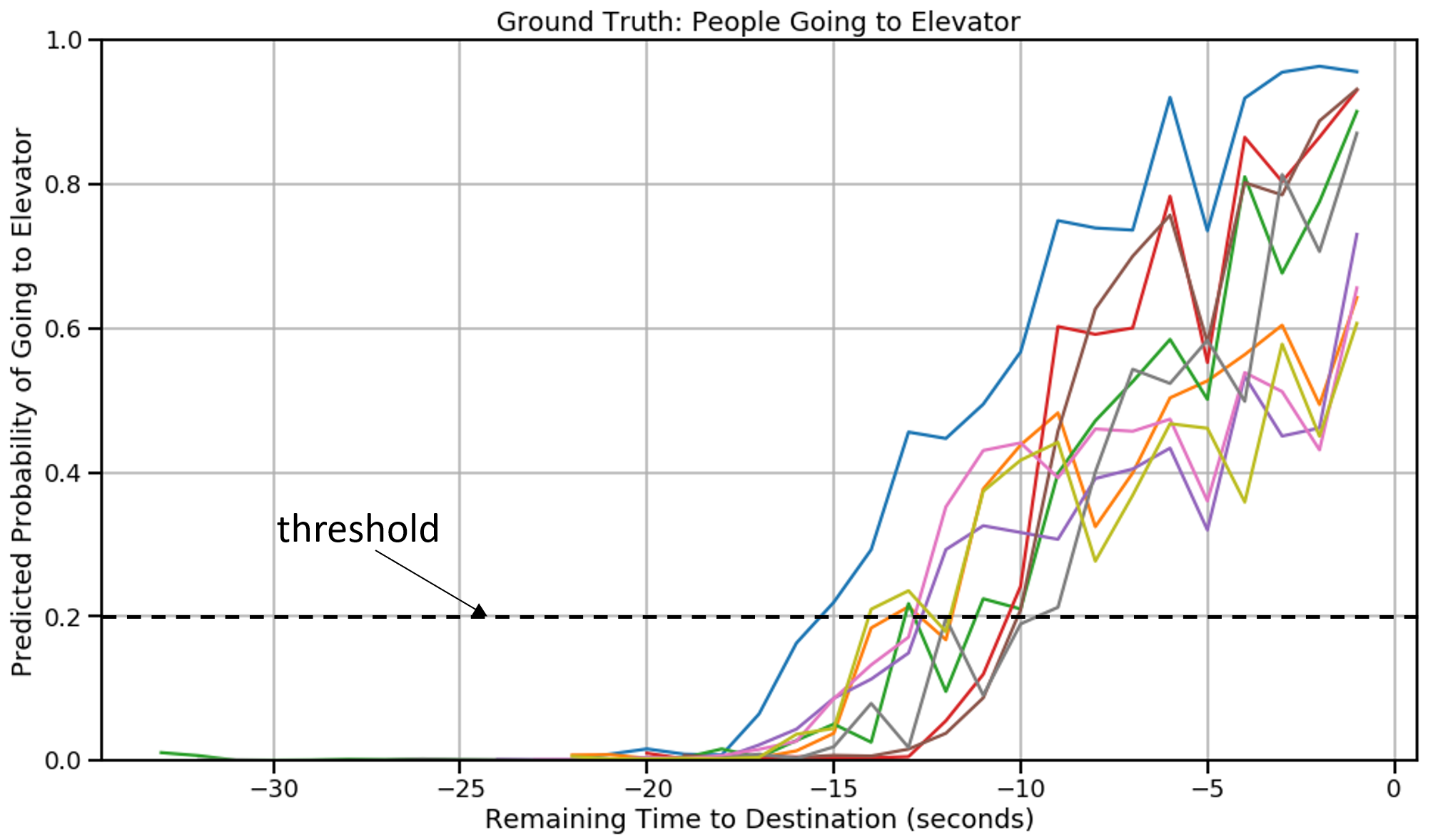}
  \caption{}
  \label{fig:ppge1}
\end{subfigure}
\caption{Predicted Probability of Going to Elevator (PPGE) vs. Remaining Time to Destination (RTD) for (a) $44$ trajectories whose ground-truth destination is NOT the elevator, and (b) $11$ trajectories whose ground-truth destination is the elevator.}
\label{fig:ppge}
\end{figure*}

	To train the Transformer model, we use the same set of parameters as that in \cite{tsiligkaridis2020personalized} for the network structure (a stack of $4$ encoder and decoder blocks), and the batch size and the number of epochs are taken as $20$ and $100$ respectively. The prediction results are shown in Fig. \ref{fig:ppge}, where we see from Fig. \ref{fig:ppge2} that for people who are not going to the elevator, the predictor will always output relatively low Predicted Probability of Going to Elevator (PPGE), in this case at most $0.112$. This suggests a simple method to eliminate practically all false positives (that is, cases when the predictor might suggest a person is going to the elevator, when in fact he/she is not) -- simply use a suitable threshold $\delta$ on the PPGE, for example $0.2$, and set the PPGE to $0$, if it is below this threshold. In practice, the threshold can be determined from the training data set, or a separate validation data set, by observing how high the PPGE gets for people who are not going to the elevator. Eliminating such low probabilities would avoid dispatching cars to pick up non-existing passengers, and at the same time, the threshold is low enough to allow detection of actual elevator passengers early enough to allow a car to be pre-dispatched on time to reduce the waiting times of such passengers. By inspecting Fig. \ref{fig:ppge1}, it can be seen that for those people who are actually going to the elevator, the destination predictor can know that with high accuracy when the Remaining Time to Destination (RTD) is as large as $10$ seconds, with only $2$ of the $11$ people whose ground-truth destination is the elevator being missed by the predictor, and when the RTD is $5$ seconds, none of these $11$ people would be missed. Note that missing future elevator landings would not change the fact that the PGES's performance is at least as good as the myopic scheduler; in particular, even if the destination predictor missed arrivals of several future passengers for the elevator, the scheduler would simply not be able to utilize the predictive information of these passengers, but it would not do worse than the myopic scheduler.

	\subsection{Accuracy of Linear Regression for Prediction of Remaining Time to Destination}
	
	From the last section, we see that the Transformer-based destination predictor is able to predict whether a moving person would head to the elevator with a high accuracy. This provides us the convenience of further predicting the remaining time to the elevator. Noting that we are actually only interested in the special destination -- the elevator, in this section we only present results from the linear regression model built for the elevator (d16). 
	
	Without loss of generality, we still use floor $\#8$ as an example to demonstrate our results. From the $11$ training trajectories with the elevator (d16) as the destination, we extract a training data set consisting of $468$ entries (e.g., an entry $((43, 40), 2.0)$ means $x^\ast = 43, y^\ast = 40, t^\ast = 2.0$ (seconds) for a point in one of the $11$ training trajectories) for the linear regression model. The test data set, also consisting of $468$ entries, is extracted from the $11$ test trajectories with the elevator (d16) as the destination. The resulting Root Mean Square Error (RMSE) of the linear regression model is $1.29$ (seconds). Considering the total travel time corresponding to a typical test trajectory is above $20$ seconds, the RMSE is satisfactorily small.
	
	It is worth pointing out that in our experiments we have also tested several more advanced regression models (random forest, decision tree regression with AdaBoost, gradient boosting regression, etc.), but their RMSE scores are very close to that of a linear regression model. This might be due to the fact that our problem is rather simple (only two explanatory variables are included) and the prediction accuracy can readily be guaranteed by using linear regression.

	\subsection{Effectiveness of the Predictive Group Elevator Scheduler}    \label{sec:accPGES}
	
	In this section, we present results from evaluating the proposed predictive group elevator scheduler that uses Transformer networks in comparison to a myopic one, as well as a predictive scheduler that uses a trivial predictor that predicts the closest location to be the final destination of the passenger. We take the prediction length $T$ to be $10$ seconds and the PPGE threshold $\delta$ to be $0.2$. We consider an elevator bank of $3$ cars for the $8$-floor building as specified in Section \ref{sec:building}.
	
	\subsubsection{Single Arrival Rate}
	
	In a single scenario experiment with a medium arrival rate ($64.8 \times 7 = 453.6$ pph, assuming floors $\#2$-$7$ also have the same arrival rate as that of floor $\#8$), we obtain the following AWT results: (1)
	the prescient scheduler (if the actual arrivals of passengers were completely known): $13.2$ seconds; (2) the Transformer-based predictive scheduler (with continuations from the Transformer based destination predictor and the linear regression based RTD predictor): $15.3$ seconds; (3) the closest-distance based predictive scheduler (the destination with the closest distance to the current location would be assigned a probability equal to $1$ of being the true final destination): $17.4$ seconds; (4) the myopic scheduler (that does not use future information): $18.1$ seconds. This results in more than $15\%$ savings in AWT when using our PGES (as opposed to the myopic scheduler). Comparatively, the rate of the AWT savings for the prescient scheduler is about $27\%$ and for the closest-distance based predictive scheduler is about $4\%$.

	\subsubsection{Variable Arrival Rates}

\begin{figure}[thpb]  
		\centering
		\includegraphics[scale=0.25]{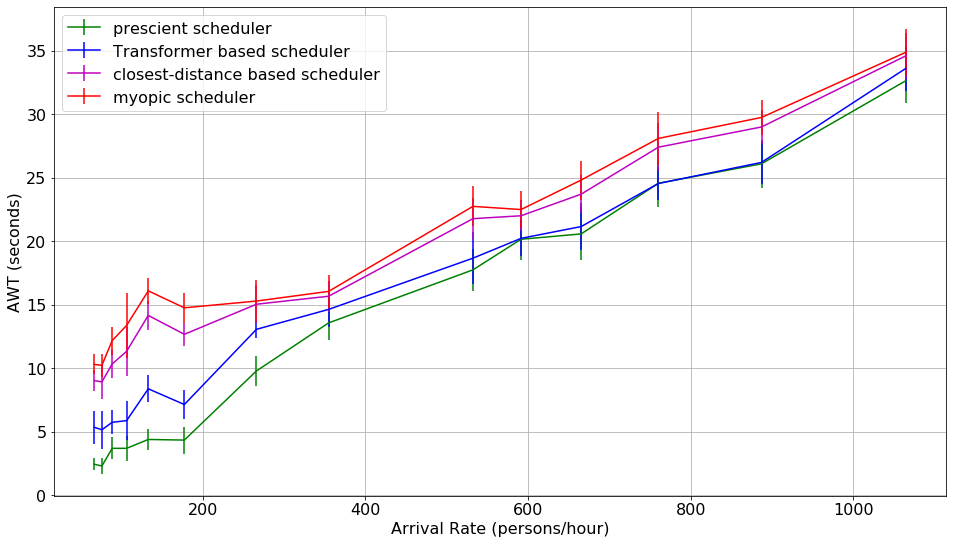}
		\caption{AWT value of four schedulers vs. arrival rate with error bars.}
		\label{fig:awt_abs}
	\end{figure}

\begin{figure}[thpb]  
		\centering
		\includegraphics[scale=0.25]{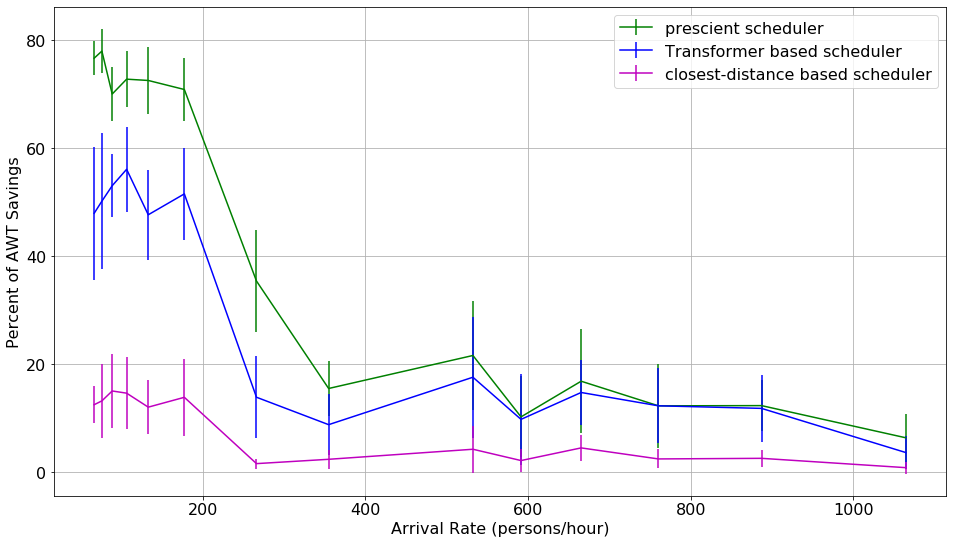}
		\caption{Percent of AWT savings of three predictive schedulers (with respect to the myopic scheduler) vs. arrival rate with error bars.}
		\label{fig:awt_imp}
	\end{figure}

	In Figures \ref{fig:awt_abs} and \ref{fig:awt_imp}, we show results of AWT values and their savings from experiments under variable arrival rates. These are obtained by averaging over 50 runs for each arrival rate, and we also show the standard deviations for these values as error bars. It is seen that the proposed Transformer-based scheduler can lead to significant AWT savings when the arrival rate is relatively small; the actual savings are around $50\%$, approaching an upper bound of above $75\%$ for the prescient scheduler. The gap could be further narrowed down, depending on the predictive performance of the Transformer/linear regression models. On the other hand, when there are denser arrivals, the PGES would have less AWT savings (e.g., when we have a medium arrival stream ($360$ pph - $720$ pph), the AWT savings of the Transformer based scheduler is around $15\%$); this is likely due to the fact that with such dense arrival streams, the bank of elevators has less capacity to leverage the available predictive information by pre-dispatching a car ahead of requests for service, and instead must focus on moving the relatively many passengers down to the lobby as efficiently as possible by matching optimally calls for service from different floors. In this regime, if all cars are busy most of the time, pre-dispatching cannot occur, but still some savings in AWT can be realized by anticipating arrivals at landings and timing the movements of cars accordingly to intercept them.
	
		From a computational complexity perspective, we note that it is unnecessary to train the Transformer model repeatedly for various arrival rates, as long as there are no significant changes in the prior probabilities for destinations.

	\section{CONCLUSION}

	In this paper, we proposed a Predictive Group Elevator Scheduler that uses predictive information about passengers arrivals from a Transformer-based destination predictor and a linear regression model that predicts remaining time to destination. The effectiveness and efficiency of our approach was validated by empirical experiments, through which we found that the savings in AWT could be as high as $50\%$ for light arrival streams, and around $15\%$ for medium arrival streams in afternoon down-peak traffic regime. These results are obtained after carefully setting the threshold of the predicted probability of going to the elevator, thus avoiding a majority of false predictions for people heading to the elevator, while achieving as high as above $80\%$ of true predictive elevator landings as early as after having seen only $60\%$ of the whole trajectory of a passenger. When such a threshold on the probability can be established, the PGES algorithm can work with a single future continuation of the arrival stream, for faster computation. The effectiveness of the PGES that uses Transformer networks for prediction in reducing the AWT is significantly higher than that of a PGES that uses a much simpler predictive method that assumes that the closest location to passengers' current location is their final destination, demonstrating the advantages of an advanced predictive method that can identify early on the likely final destination.
	
    For future work, we will consider building predictive schedulers for various periods of the day (hence, various prior probabilities of the passengers' destinations). One possibility is training separate models for each period, and another one is to employ transfer learning approaches between different periods. Another promising avenue for improving prediction accuracy is to customize predictive models for each individual passenger (in a privacy preserving manner), opening the possibility for truly smart buildings that are highly adaptive and responsive to the habits and schedules of all of their individual occupants.

	\addtolength{\textheight}{-12cm}   % This command serves to balance the column lengths
	
	%%%%%%%%%%%%%%%%%%%%%%%%%%%%%%%%%%%%%%%%%%%%%%%%%%%%%%%%%%%%%%%%%%%%%%%%%%%%%%%%

	%%%%%%%%%%%%%%%%%%%%%%%%%%%%%%%%%%%%%%%%%%%%%%%%%%%%%%%%%%%%%%%%%%%%%%%%%%%%%%%%

	%%%%%%%%%%%%%%%%%%%%%%%%%%%%%%%%%%%%%%%%%%%%%%%%%%%%%%%%%%%%%%%%%%%%%%%%%%%%%%%%

	%\section*{ACKNOWLEDGMENT}

	%%%%%%%%%%%%%%%%%%%%%%%%%%%%%%%%%%%%%%%%%%%%%%%%%%%%%%%%%%%%%%%%%%%%%%%%%%%%%%%%

	%\begin{thebibliography}{99}
	\bibliographystyle{IEEEtran} % to show URLs in references in the working draft of the paper
	\bibliography{DanielsReferencesFromMendeley-old}  % .bib
	%\end{thebibliography}

\end{document}